\documentclass[entropy,review,accept,moreauthors,pdflatex,12pt,a4paper]{mdpi}

\usepackage{graphicx}

\usepackage{bm}
\usepackage{epsfig}
\usepackage[colorlinks=true,linkcolor=blue]{hyperref}

\lastpage{38}
\doinum{10.3390/------}
\pubvolume{18}
\pubyear{2016}
\history{Received: November 28, 2015 / Accepted: March 3, 2016 / Published: xx}

\newcommand{\be}{\begin{equation}}
\newcommand{\ee}{\end{equation}}

\newcommand{\bea}{\begin{eqnarray}}
\newcommand{\eea}{\end{eqnarray}}

\newcommand{\al}{\alpha}

\newcommand{\gm}{\gamma}
\newcommand{\Gm}{\Gamma}

\newcommand{\eps}{\epsilon}

\newcommand{\et}{\eta}

\newcommand{\kp}{\kappa}

\newcommand{\Lm}{\Lambda}

\newcommand{\rh}{\rho}

\newcommand{\sg}{\sigma}

\newcommand{\Om}{\Omega}

\newcommand{\fdot}{\mbox{\boldmath $\cdot$}}

\newcommand{\rarrow}{\rightarrow}

\newcommand{\nn}{\nonumber}

\newcommand{\varep}{\varepsilon}

\Title{Dark Energy: The Shadowy Reflection of Dark~Matter?}

\Author{Kostas Kleidis $^{1,}$ and Nikolaos K. Spyrou $^2$}

\address{$^{1}$ Department of Mechanical Engineering, Technological Education Institute of Central Macedonia, Serres~621.24, Greece; kleidis@teiser.gr \\
$^{2}$ Department of Astronomy, Aristoteleion University of Thessaloniki, Thessaloniki~541.24, Greece; spyrou@auth.gr}  

\corres{Kostas Kleidis}

\abstract{In this article, we review a series of recent theoretical results regarding a conventional approach to the dark energy (DE) concept. This approach is distinguished among others for its simplicity and its physical relevance. By compromising General Relativity (GR) and Thermodynamics at cosmological scale, we end up with a model without DE. Instead, the Universe we are proposing is filled with a perfect fluid of self-interacting dark matter (DM), the volume elements of which perform hydrodynamic flows. To the best of our knowledge, it is the first time in a cosmological framework that the energy of the cosmic fluid internal motions is also taken into account as a source of the universal gravitational field. As we demonstrate, this form of energy may compensate for the DE needed to compromise spatial flatness, while, depending on the particular type of thermodynamic processes occurring in the interior of the DM fluid (isothermal or polytropic), the Universe depicts itself as either decelerating or accelerating (respectively). In~both cases, there is no disagreement between observations and the theoretical prediction of the distant supernovae (SNe) Type Ia distribution. In fact, the cosmological model with matter content in the form of a thermodynamically-involved DM fluid not only interprets the observational data associated with the recent history of Universe expansion, but also confronts successfully with every major cosmological issue (such as the age and the coincidence problems). In this way, depending on the type of thermodynamic processes in it, such a model may serve either for a conventional DE cosmology or for a viable alternative one.}



\begin{document}

\section{Introduction}

During the last 20 years, a continuously growing list of observational data has verified the existence of a distributed energy component in the Universe, \emph{i.e.}, one that does not seem to cluster at any scale. This new constituent of the cosmic matter-energy content was termed dark energy \cite{1,2}; a reflection of our ignorance on its exact nature, the determination of which has become one of the biggest problems in theoretical physics and cosmology (for a review of the various DE models see, e.g., \cite{3}). Let us briefly review how did we get to this point.

It all begun in the late 1990s, when high-precision distance measurements revealed that, in a dust Universe, the far-off light-emitting sources look fainter (in other words, their actual distance is larger) than what is theoretically predicted \cite{4,5,6,7,8,9,10,11,12,13,14,15,16,17,18,19,20,21,22,23,24,25,26,27,28,29,30}. To compromise theory with observation, \mbox{Perlmutter \emph{et al}. \cite{2}} and Riess \emph{et al}. \cite{9} admitted that the value of the long-sought cosmological constant, $\Lm$, is no longer zero (in connection, see \cite{31}). In this case, along with its mass content, the Universe is filled also with an extra, uniformly distributed amount of energy.

At the same time, detailed observational studies of the cosmic microwave background (CMB) suggested that the post-recombination Universe can be described (to high accuracy) by a spatially-flat Robertson-Walker (RW) model \cite{32,33,34,35,36,37,38,39,40}. This means that the overall energy density, $\varep$, of the Universe matter-energy content, in units of the critical energy density, $\varep_c = \rh_c c^2$ (the equivalent to the critical rest-mass density, $\rh_c= \frac{3 H_0^2}{8 \pi G}$, where $H_0$ is the Hubble parameter at the present epoch, $G$ is Newton's gravitational constant, and $c$ is the velocity of light), must be very close to unity, $\Om = \frac{\varep}{\varep_c} \simeq 1$, \emph{i.e.}, much larger than the (most recently) measured value of the density parameter, $\Om_M = \frac{\rh}{\rh_c} = 0.279$ \cite{41}. Once again, an extra amount of energy was needed, this time, in order to compromise spatial flatness.

The particle-physics vacuum does contribute a uniformly distributed energy component, which could serve as an effective cosmological constant and justify spatial flatness \cite{42}. Unfortunately, vacuum energy is $10^{123}$ times larger than what is currently inferred by observations (see, e.g., \cite{43}). Clearly, to reconcile spatial flatness with the observed dimming of the cosmological standard candles, another approach, \emph{i.e.}, other than the cosmological constant, was needed.

As a consequence, several physically-motivated models have appeared in the literature, such as, models with scalar fields \cite{44,45}, phantom cosmology \cite{46}, tachyonic matter \cite{47}, braneworld scenarios \cite{48,49}, scalar-tensor theories \cite{50}, $f(R)$-gravity \cite{51}, holographic gravity~\mbox{\cite{52,53,54}}, Chaplygin gas \cite{55,56,57,58}, Cardassian cosmology \cite{59,60,61}, models with extra (\emph{i.e.}, more than four) dimensions \cite{62,63,64,65}, neutrinos of varying mass \cite{66,67}, and many others (see, e.g., \cite{68}). However, most of these models are suffering by the (old) cosmological-coincidence problem. According to it, a viable DE model should be able (also) to explain why we live so close to the transition era; the inflection point being (observationally) set at a rather low value of the cosmological redshift parameter, $z$, the so-called transition redshift, $z_{tr} = 0.752 \pm 0.041$ \cite{30}. Of course, we should mention that, it is not completely clear how much the transision redshift depends (or not) either on the specific cosmological model used \cite{69,70} or on the particular theory of gravitation that is taken into account~\cite{71,72,73}. In other words, for the time being, $z_{tr}$ cannot be measured in a model-independent way; and this is where, usually, cosmography takes over \cite{74}.

In view of the cosmographic approach, all quantities of interest are expanded as Taylor series around their present-time values, with the corresponding coefficients being directly related to several parameters of cosmological significance \cite{75}. In other words, cosmography is a technique for matching cosmological data with observable quantities without imposing a particular cosmological model. In view of the large number of speculative models presently being considered, such an observationally-driven approach is of interest in its own right; hence, many cosmographic efforts to resolve the controversy on the exact nature of the cosmic fluid have appeared in the literature (see, e.g., \cite{76,77,78,79,80,81,82,83,84,85,86,87,88,89,90,91,92,93,94,95,96,97}).

In the meantime, much evidence in favour of a dark (energy) component in the Universe matter-energy content had been accumulated, also from observations of galaxy clusters \cite{98}, the integrated Sachs-Wolfe (ISW) effect \cite{99}, baryon acoustic oscillations (BAOs) \cite{100,101}, weak gravitational lensing (WGL) \cite{102,103}, and the Lyman-$\al$ (LYA) forest \cite{104}. For the first time since the early 1930s, observation was prevailing over theory. As a consequence, many alternative interpretations to the DE concept also appeared in the literature (see, e.g., \cite{105,106,107,108}), although a dark component was already present in the Universe matter content.

Indeed, nowadays, there is too much evidence in favour of a non-baryonic mass component in the Universe matter content. This evidence includes high-precicion measurements of the flattened galactic rotation curves \cite{109,110}, the WGL of distant galaxies by (some dark) foreground structure~\cite{111}, and the weak modulation of strong lensing around individual massive elliptical galaxies \cite{112}. On the scale of galaxies, recent observational data indicate that the (dark) galactic haloes extend almost half the distance to the neighboring galaxies \cite{113,114}. On larger scales, it has been found that the total-mass of galaxy clusters is almost ten times higher than its baryonic counterpart \cite{115,116,117}, while, analogous conclusions can be deduced also at the universal level, from the combination of CMB measurements \cite{39} with those concerning light-chemicals abundances \cite{118}. Accordingly, more than $85 \%$ of the mass in the Universe consists of non-luminous DM \cite{119}.

The precise nature of the DM constituents is still a matter of debate. One of the most attractive candidates are the weakly interacting massive particles (WIMPs)---a by-product of the Universe hot youth 
\cite{120,121,122}. These particles are quite relevant to the direct or/and indirect detection of DM, due to their connection to standard-model particles \cite{123,124,125}. However, for such a candidancy only weak-scale physics is involved, and, therefore, cosmologists used to argue that the WIMPs are practically collisionless. However, recent results from high-energy particle detectors \cite{126,127}, combined with data from the Wilkinson Microwave Anisotropy Probe (WMAP) \cite{128}, have revealed an unexplained positron excess in the Universe, which could be related to interactions between DM particles (see, e.g., \cite{129,130,131,132,133,134,135,136,137,138,139}). In other words, the WIMPs can be slightly collisional \cite{140,141,142,143,144}. 

A collisional-DM model could reconcile DM and DE in terms of a single component, thus arising as a relatively inexpensive solution to the DE problem (see, e.g., \cite{145,146,147,148,149,150,151,152,153,154,155,156,157,158}). In this framework, Kleidis and Spyrou \cite{159,160}, proposed a classical approach to the DE concept, in terms of a phenomenological model, in which the Universe matter content possesses (also) some sort of thermodynamic properties. Indeed, in view of the CMB-based spatial flatness, today, the Universe should contain a much larger amount of energy than what is attributed to the total rest-mass of its matter content. This, however, would have no longer been a problem, if the dominant component of the Universe matter content (\emph{i.e.}, DM) was represented by a thermodynamically-involved fluid, in which, the extra energy needed to compromise spatial flatness is attributed to the energy of its internal motions. Notice that, the same assumption has been proved very useful in modeling dark galactic haloes, leading to a significant improvement of the galaxies velocity dispersion profiles \cite{161,162,163,164,165,166,167}. 

It is therefore worth examining the properties of a cosmological model, in which, in principle, there is no DE at all. Instead, we assume that the evolution of this model is driven by collisional-DM, \emph{i.e.}, a cosmic fluid with thermodynamical content. In this case, the fundamental constituents of the Universe matter content are the volume elements of this fluid, performing hydrodynamic flows. We~distinguish two types of thermodynamic processes in it, namely, (i) isothermal flows \cite{159} and (ii) polytropic flows \cite{160}. In both cases, the energy of the DM fluid internal motions is also taken into account as a source of the universal gravitational field, thus compensating for the extra DE needed to compromise spatial flatness. In such a cosmological model, there is no disagreement between the theoretical prediction and the observed distrubution of the distant SNe Ia, while neither the age problem nor the corresponding coincidence one ever rise.

This review article is organized as follows: In Section \ref{sec2}, we consider a spatially-flat Universe, the evolution of which is driven by a (perfect) fluid of self-interacting (\emph{i.e.}, thermodynamically-involved) DM. Accordingly, in Section \ref{sec3}, we focus on the special case where the volume elements of this fluid perform isothermal flows \cite{159}. In this case, an extra DE amount arises naturally---being represented by the energy of the internal motions of the DM fluid---although the Universe is ever-decelerating. However, what we really need to query about is, what is realized by an observer who - although living in such a model---mistreats DM as collisionless dust. As we shall demonstrate, for such an observer, besides the need for a DE amount (to compromise spatial flatness), every cosmologically-distant indicator appears to be fainter (\emph{i.e.}, its actual distance is larger) than what is theoretically predicted, and the late Universe is accelerating. Although intriguing, this alternative model has a delicate point: It is compatible with observations, only if the matter content of the dark sector consists of hot DM (HDM). However, pure HDM models can not reproduce the observed large-scale structure of the Universe \cite{168}, in contrast to their cold DM (CDM) counterparts \cite{169}. In an effort to confront with this issue, in Section \ref{sec4}, we consider that the dominant type of process occuring in the interior of the (cosmic) DM fluid is polytropic flow \cite{160}. Once again, the extra DE needed to compromise spatial flatness is represented by the internal energy of the fluid. The polytropic (DM) model depends on one free parameter, the polytropic exponent, $\Gm$. For $\Gm \leq 0.540$ the pressure of the cosmic fluid is definitely negative, and the Universe does accelerate its expansion, below a transition value of the cosmological redshift, $z_{tr}$. In fact, the polytropic DM model can confront with almost every major cosmological issue, such as the age and the coincidence problems, while reproducing to high accuracy the observed distribution of the SNe Ia standard candles. Finally, we conclude in Section \ref{sec5}.

\section{Collisional-DM Cosmology}\label{sec2}

In view of the CMB observational data released at the dawn of the 21st century (see, e.g., \cite{32,33,34,35,36}), the post-recombination Universe is described by a spatially-flat RW model, the line element of which is given by \be ds^2 = S^2 (\et) \left [ c^2 d \et^2 - \left ( dx^2 + dy^2 + dz^2 \right ) \right ] \: , \label{eq1}\ee where $\et$ is conformal time, and $S (\et)$ is the scale factor. As a consequence, the value of the Hubble parameter at the present epoch is, by definition, given by \be H_0^2 = \frac{8 \pi G}{3} \rh_c \label{eq2}\ee (see, e.g., \cite{170} (p. 77)). The evolution of model (\ref{eq1}) depends on the nature of the source that drives the universal gravitational field, \emph{i.e.}, its matter-energy content. 

According to Kleidis and Spyrou \cite{159,160}, along the lines of the collisional-DM approach, we assume that, in principle, there is no DE at all. Instead, we admit that the DM possesses fluid-like properties, in the sense that, the collisions of WIMPs maintain a tight coupling between these particles, so that their kinetic energy is re-distributed. Under this assumption, the DM acquires some sort of thermodynamical content, and, therefore, the evolution of the post-recombination Universe is no longer driven by pressureless dust, but by a fluid, which, in view of the cosmological principle, should (rather) be practically homogeneous and isotropic at large scale. The pressure of this (perfect) fluid is accordingly given by a barotropic equation of state, \be p = f (\rh)  \: , \label{eq3}\ee where $\rh$ is the rest-mass density, \emph{i.e.}, the part equivalent to the energy density $\rh c^2$ that remains unaffected by the internal motions of the cosmic fluid. Now, the fundamental units of the Universe matter content are the volume elements of the collisional-DM fluid (elements of fluid, each one consisting always of the same number of particles).

In the context of GR, the motions of volume elements in the interior of a continuous medium are governed by the equations \be T_{\; \; ; \nu}^{\mu \nu} = 0 \; , \label{eq4}\ee where Greek indices refer to the four-dimensional spacetime, Latin indices refer to the three-dimensional spatial slices, the semicolon denotes covariant derivative, and $T^{\mu \nu}$ is the energy-momentum tensor of the Universe matter content. In the particular case of a perfect fluid, $T^{\mu \nu}$ admits the standard form \be T^{\mu \nu} = (\varep + p)u^{\mu} u^{\nu} - p g^{\mu \nu} \: , \label{eq5}\ee where $u^{\mu} = \frac{dx^{\mu}}{ds}$ is the four-velocity $\left ( u_{\mu}u^{\mu} = 1 \right )$ at the position, $x^{\mu}$, of a fluid volume element, $g^{\mu \nu}$ are the contravariant components of the Universe metric tensor, and $\varep$ is the total energy density of the fluid. In an ideal equilibrium state (\emph{i.e.}, in the absence of shear and viscocity), $\varep$ is decomposed to \be \varep = {\cal E} (\rh, T) + \rh \: {\cal U} (T) \label{eq6} \ee (see, e.g., \cite{171} (pp. 81--84 and 90--94)). In Equation~(\ref{eq6}), $T$ is the absolute temperature and ${\cal U}$ is the energy of this fluid internal motions, thus defining $\rh \: {\cal U}$ as the corresponding (specific) energy density. In this framework, ${\cal E} (\rh, T)$ represents every form of energy (density) involved, other than that of internal motions (e.g., due to the rest-mass content, heat exchange, \emph{etc}.). Along these lines, Equation~(\ref{eq4}) represent the hydrodynamic flows of the volume elements in the interior of a perfect-fluid source.

However, in a maximally symmetric cosmological setup, an observer comoving with the cosmic expansion also traces the hydrodynamic flow of the spatially-homogeneous cosmic fluid and the Weyl postulate is valid (see, e.g., \cite{172} (p. 91)). Accordingly, the dynamical evolution of the cosmological model with line element given by Equation (\ref{eq1}) is governed by the equation of the classical (\emph{i.e.}, for $\Lm = 0$) Friedmann-Robertson-Walker (FRW) cosmology \be H^2 = \frac{8 \pi G}{3 c^2} \varep \: ,\label{eq7} \ee where \be H = \frac{S^{\prime}}{S^2} \label{eq8} \ee is the Hubble parameter as a function of the scale factor, and the prime denotes differentiation with respect to $\et$. As regards Equation~(\ref{eq7}), we need to stress that there is an essential difference between our (thermodynamical) model and the rest of the classical FRW cosmologies, since, in this case, the basic matter constituents are no longer particles receding from each other, but the volume elements of the collisional-DM fluid, which possess some sort of internal structure and, hence, thermodynamical content. Therefore, the functional form of $\varep$ in Equation~(\ref{eq7}) is no longer given by $\rh c^2$ alone, but by Equation~(\ref{eq6}) (in connection, see also \cite{172} (pp. 61--62)). Accordingly, we need to determine ${\cal E}$ and ${\cal U}$. To do so, we address (i) to the first law of thermodynamics in curved spacetime, \be d \: {\cal U} + p d \left ( \frac{1}{\rh} \right ) = {\cal C} d T \label{eq9}\ee (see, e.g., \cite{171} (p. 83)), where ${\cal C}$ is the specific heat of the cosmic fluid, and (ii) to the conservation law of GR, $T_{\; ; \nu}^{0 \nu} = 0$, which, in terms of the metric tensor, $g_{\mu \nu}$, associated to Equation~(\ref{eq1}), results in the continuity equation \be \varep^{\prime} + 3 \frac{S^{ \prime}}{S} (\varep + p) = 0 \: . \label{eq10}\ee 

Now, in order to proceed further, we need to decide on the type of processes that take place in the interior of the cosmic (DM) fluid, \emph{i.e.}, to determine the functional form of the equation of state given by Equation~(\ref{eq3}). To do so, we distinguish two cases, namely, (i) isothermal flows \cite{159} and (ii) polytropic flows \cite{160}. We consider each one of these cases, separately.

\section{Isothermal Processes in a Cosmological DM Fluid}\label{sec3}

In the case where the volume elements of the cosmic fluid perform isothermal flows, we have $dT = 0 = dQ$, and, hence, ${\cal E} = \rh c^2$. Accordingly, the DM (always entangled with the small baryonic contamination) constitutes a gravitating perfect fluid with equation of state \be p = w \rh c^2  \: , \label{eq11}\ee where $0 \leq w = \left ( \frac{c_s}{c} \right )^2 \leq 1$ is a dimensionless constant, which measures the square of the speed of sound, $c_s$, in units of $c^2$. For $dT = 0$, the first law of Thermodynamics Equation~(\ref{eq13}) results in \cite{173} \be {\cal U} = {\cal U}_0 + wc^2 \ln \left ( \frac{\rh}{\rh_0} \right ) \: , \label{eq12}\ee where $\rh_0$ and ${\cal U}_0$ are the present-time values of rest-mass density and internal-energy density, respectively. By virtue of Equation~(\ref{eq12}), the total energy density of the Universe matter-energy content is written in the form \be \varep = \rh c^2 \left [ 1 + \frac{{\cal U}_0}{c^2} + w \ln \left ( \frac{\rh}{\rh_0} \right ) \right ] \label{eq13}\ee and the continuity Equation (\ref{eq10}) yields \be \rh = \rh_0 \left ( \frac{S_0}{S} \right )^3 \: ,\label{eq14} \ee where $S_0$ is the present-time value of the scale factor. Accordingly, in the isothermal case, the Friedmann Equation (\ref{eq7}) reads \be \left ( \frac{H}{H_0} \right )^2 = \Om_M \left ( \frac{S_0}{S} \right )^3 \left [ 1 + \frac{{\cal U}_0}{c^2} + 3 w \ln \left ( \frac{S_0}{S} \right ) \right ] \: , \label{eq15}\ee where we have used also Equation~(\ref{eq2}). At the present epoch (when $S = S_0$ and $H = H_0$), Equation~(\ref{eq15}) is reduced to \be {\cal U}_0 = \frac{1 - \Om_M}{\Om_M} c^2 \label{eq16}\ee and, therefore, from Equation~(\ref{eq6}) we find that, in a cosmological model with matter content in the form of an isothermal DM fluid, the present-time value of the overall-energy density parameter is exactly unity, \emph{i.e.}, \be \Om_0 =  \frac{\varep_0}{\varep_c} = \frac{\rh_0 c^2}{\rh_c c^2} + \frac{\rh_0 {\cal U}_0}{\rh_c c^2} = \Om_M + \Om_M \frac{{\cal U}_0}{c^2} = 1 \: . \label{eq17}\ee 

In other words, the (extra) DE needed to flatten the Universe can be represented by the energy of the internal motions of the isothermal DM fluid (iDMF model).

By virtue of the present-time value of the internal energy given by Equation~(\ref{eq16}), Equation~(\ref{eq15}) results in \be \left ( \frac{H}{H_0} \right )^2 = \left ( \frac{S_0}{S} \right )^3 \left [ 1 + 3 w \Om_M \ln \left ( \frac{S_0}{S} \right ) \right ] \: . \label{eq18}\ee 

The Friedmann Equation (\ref{eq18}) that determines the evolution of the iDMF model, can become very useful, if we take into account that, since $0 \leq w \leq 1$ and $\Om_M = 0.274$ \cite{40}, the combination $w \Om_M$ can be quite small, \emph{i.e.}, $w \Om_M \ll 1$. Accordingly, to terms linear in $w \Om_M$, Equation~(\ref{eq18}) is written in the~form \be H \simeq H_0 \left ( \frac{S_0}{S} \right )^{\frac{3}{2} (1 + w \Om_M)} \: , \label{eq19}\ee which, by virtue of Equation~(\ref{eq8}), can be solved analyically, to determine the scale factor of the iDMF model, as follows \be S = S_0 \left ( \frac{\et}{\et_0} \right )^{\frac{2}{1 + 3 w \Om_M}} , \label{eq20}\ee where \be \et_0 = \frac{2}{(1 + 3 w \Om_M) H_0 S_0} \ee is the present-time value of the conformal time. For $w \neq 0$, 
Equation~(\ref{eq20}) is the natural generalization of the Einstein-de Sitter (EdS) model $\left ( S \sim \et^2 \right )$ (see, e.g., \cite{170} (pp. 77, 83 and 142--144)), \emph{i.e.}, of the collisionless-DM counterpart of the iDMF model. 

Now, upon consideration of the cosmological redshift parameter, \be z + 1 = \frac{S_0}{S} \: , \ee 

Equation~(\ref{eq19}) can be cast in the form \be H = H_0 ( 1 + z )^{\frac{3}{2} (1 + w \Om_M)} \: . \label{eq23}\ee 

In view of Equation~(\ref{eq23}), on the approach to the present epoch (when $z = 0$), $H(z)$ decreases monotonically. In other words, the cosmological model filled with an isothermal DM fluid decelerates its expansion. Indeed, in the iDMF model the deceleration parameter, \be q (z) = \frac{dH / dz}{H(z)} (1+z) - 1\label{eq24}\ee (\emph{cf.} Equation~(16) of \cite{174}), yields \be q (z) = \frac{1}{2} (1 + 3 w \Om_M) > 0 \: , \label{eq25}\ee independently of $z$, even if $w = 0$. In other words, the iDMF model cannot confront with the apparent accelerated expansion of the Universe. The actual reason is that, it does not have to. As we shall demonstrate in Section 3.1, in a cosmological model with matter content in the form of isothermal (DM) fluid, both the observed dimming of the distant SNe Ia and the apparent accelerated expansion of the Universe can be due to the misinterpretation of several cosmological parameters, by those observers who, although living in the iDMF model, insist on adopting the collisionless-DM approach.

\subsection{Mistreating DM as Collisionless}

In the late 1990s, the scientific community was assured that the Universe is filled with collisionless DM, \emph{i.e.}, mainly, dust. Then, high-precision distance measurements performed with the aid of SNe Ia events, revealed the unexpected acceleration of the cosmic expansion. However, the physical content of a dust Universe is entirely different from that of a collisional-DM model. In other words, the dynamical properties of a pressureless Universe are no longer described by the metric tensor $g_{\mu \nu}$, associated to Equation~(\ref{eq1}), but, rather, in terms of another metric tensor, $\tilde{g}_{\mu \nu}$. Clearly, someone who (mis)treats DM as dust, relies on $\tilde{g}_{\mu \nu}$ in interpreting observations. For such an observer, the accumulated evidence in favour of spatial flatness, not only implies that the spacetime line element is written in the form \be d \tilde{s}^2 = \tilde{S}^2 (\et) \left [ c^2 d \et^2 - \left ( dx^2 + dy^2 + dz^2 \right ) \right ] \: , \ee but, also, suggests that there is a deficit $(1 - \Om_M)$ in the universal energy budget. On the contrary, in the iDMF approach, the extra energy amount needed to compromise $\Om_0 = 1$ is already included in the model, being represented by the energy of the internal motions of the thermodynamically-involved DM fluid. Furthermore, on interpreting the dimming of the cosmologically-distant SNe Ia events, a supporter of the collisionless-DM scenario is also based on $\tilde{g}_{\mu \nu}$ and the cosmological parameters arising from it, hence, a possible explanation for such a dimming could be that, recently, the Universe accelerated its expansion \cite{7,9}. We cannot help but wondering, what would be the explanation within the context of the iDMF model.

In search of such an explanation, we note that, according to Kleidis and Spyrou \cite{175} the collisional-DM treatment of the Universe matter-energy content (in terms of which, in principle, $p \neq 0$) can be related to the collisionless-DM approach (in terms of which, necessarily, $\tilde{p} = 0$) by a conformal transformation, \be \tilde{g}_{\mu \nu} = f^2 (x^{\kp}) \: g_{\mu \nu} \label{eq27}\ee (see also \cite{176,177} (pp. 24--29 and 54--61), \cite{178,179}), where, upon consideration of isentropic flows, the conformal factor, $f (x^{\kp})$, is given by \cite{175} \be f (x^{\kp}) = C \left ( \frac{\varep + p}{\rh c^2} \right )\label{eq28} \ee corresponding to the specific enthalpy of the ideal fluid under consideration ($C$ is an integration constant). More recently, Verozub showed that Equations~(\ref{eq27}) and (\ref{eq28}) apply to every Riemannian spacetime, and not just to the metric tensor associated with a perfect fluid source \cite{180}.

With the aid of Equations~(\ref{eq27}) and (\ref{eq28}), we can now determine the scale factor $\tilde{S} (\et)$, \emph{i.e.}, the scale factor of the Universe as it is realized by someone who, although living in the iDMF model, mistreats DM as collisionless. By virtue of Equation~(\ref{eq27}), \be \tilde{S} (\et) = f (x^{\kp}) S (\et) \: , \label{eq29}\ee where, in terms of $z$, $f (x^{\kp})$ is given by \cite{159} \be f(z) = 1 + w \Om_M \left [ 1 + 3 \ln (1 + z) \right ] \: . \label{eq30}\ee 

Upon consideration of Equations~(\ref{eq29}) and (\ref{eq30}), we can express several cosmological parameters of the iDMF model, in terms of their collisionless-DM counterparts. In this framework, the cosmological redshift parameter, $\tilde{z}$, as defined by a supporter of the collisionless-DM scenario, is given by \be \tilde{z} + 1 = \frac{\tilde{S} (\et_0)}{\tilde{S} (\et)} = \frac{1 + w \Om_M}{1 + w \Om_M [1 + 3 \ln (1 + z)]} (z + 1) \: , \ee which, to linear terms in $w \Om_M \ll 1$, results in \be 1 + \tilde{z} \simeq (1 + z)^{1 - 3w \Om_M} \: . \label{eq32}\ee 

Equation~(\ref{eq32}) suggests that, for every value of $z$, \emph{i.e.}, the cosmological redshift as it is defined in the iDMF model, the corresponding collisionless-DM quantity, $\tilde{z}$, is always a little bit smaller $(\tilde{z} < z)$. In other words, on observing a standard candle in the iDMF model, an observer who adopts the collisionless-DM approach infers that it lies farther $(z)$ than expected $(\tilde{z})$.

\subsection{Accomodating the Recent SNe Ia Data in a Decelerating Universe}

One of the most reliable methods to monitor the Universe expansion, is to measure the redshift and the apparent magnitude, $m$, of cosmologically-distant SNe Ia events (standard candles), whose absolute magnitude, $M$, is well-known \cite{5,6,7,8,9}. In an effort to determine the distribution of these events in curved spacetime, a number of scientific groups found evidence in favour of a recent accelerating stage of the Universe expansion \cite{6,7,8,9,10,11,12,13,14,15,16,17,18,19,20,21,22,23,24,25,26,27,28,29,30}. Indeed, in all of these surveys, the SNe Ia events (at peak luminocity) look fainter (\emph{i.e.}, their actual distance is larger) than what is theoretically predicted. This~result led the scientific community to admit that, recently, the Universe (driven by an exotic DE fluid of negative pressure) accelerated its expansion (see, e.g., \cite{37}). However, in view of Equation~(\ref{eq32}), there may be another, more conventional interpretation. 

To begin with, we note that photons travel along null geodesics, $d \tilde{s}^2 = 0 = ds^2$, which remain unaffected by conformal transformations. Consequently, both in the collisional-DM treatment and in the collisionless-DM approach, the radial distance of a particular SN Ia event is the same, \emph{i.e.}, $\tilde{r} = r$. In this case, upon consideration of Equation~(\ref{eq32}), the actually-measured (in a spatially-flat iDMF model) luminosity distance, \be d_L (z) = r S(\et_0) (1 + z) \: , \ee can be expressed in terms of the corresponding collisionless-DM quantity, \be \tilde{d}_L (\tilde{z}) = \tilde{r} \tilde{S} (\et_0) (1 + \tilde{z}) \: , \ee as follows \be \frac{d_L }{\tilde{d}_L} = \frac{1}{1 + w\Om_M} ( 1 + z )^{3 w \Om_M} \: . \label{eq35}\ee 

According to Equation~(\ref{eq35}), in the iDMF Universe, there exists a characteristic value of the actually-measured cosmological redshift, namely, \be z_c = \left ( 1 + w \Om_M \right )^{\frac{1}{3w \Om_M}} - 1 \: , \label{eq36}\ee such that, for $z > z_c$, $d_L > \tilde{d}_L$. In other words, an observer who---although living in the iDMF model---treats DM as dust, infers that any SN Ia event located at $z > z_c$ lies farther than what is theoretically predicted. By virtue of $z_c$, an inflection point (on the $d_L$ \emph{versus} $z$ diagram) arises naturally in the iDMF model, without the need to assume any transition from deceleration to acceleration. In~other words, the iDMF model does not suffer from the coincidence problem.  

Now, we shall demonstrate that, in fact, there is no discrepancy between theoretical prediction and the observed distribution of the distant SNe Ia in the iDMF model. To do so, we overplot the theoretically-determined distance modulus (of a light-emitting source in the iDMF model) on the Hubble ($\mu$ \emph{versus} $z$) diagram of the sample of 192 SN Ia events used by \mbox{Davis \emph{et al}. \cite{181}} 
(Available at http://braeburn.pha.jhu.edu/$\sim$ariess/R06). In what follows, we admit that \mbox{$H_0 = 70.2 \: Km/sec/Mpc$~\cite{40}} and hence $2 c/H_0 = 8,547 \: Mpc$. 

In the iDMF model, the distance modulus of a cosmologically-distant indicator is given by \be \mu (z) = 5 \log \left [ \frac{d_L (z)}{Mpc} \right ] + 25 = m (z) - M \: , \label{eq37}\ee where $d_L$ is measured in megaparsecs $(Mpc)$. In a similar manner, \be \tilde{\mu} (\tilde{z}) = 5 \log \left [ \frac{ \tilde{d}_L (\tilde{z})}{Mpc} \right ] + 25 \label{eq38}\ee is the theoretical formula associated to the distance modulus of the same source, as it is defined by someone who, although living in the iDMF model, insists on adopting the (traditional) collisionless-DM approach. In this case, using Equation~(\ref{eq36}), we obtain \be \mu - \tilde{\mu} = 15 w \Om_M \log ( 1 + z ) - 5 \log \left ( 1 + w \Om_M \right ) \: . \label{eq39}\ee 

According to Equation~(\ref{eq39}), any light-emitting source of the iDMF model that is located at $z > z_c$, from the point of view of an observer who treats DM as collisionless, appears to be dimmer than expected, \emph{i.e.}, $\tilde{\mu} < \mu$. Therefore, if the cosmic DM amount is interpreted as an isothermal fluid, then, it is possible that the discrepancy between the expected value of the distance modulus $(\tilde{\mu})$ of a distant SN Ia and the corresponding observed one $(\mu)$, arises only because some cosmologists (although living in the iDMF model) insist on treating DM as dust. 

In order to overplot Equation~(\ref{eq38}) on the $\mu$ \emph{versus} $z$ diagram of the SN dataset used by \mbox{Davis \emph{et~al}.~\cite{181},} first, we need to determine the luminosity distance $\tilde{d}_L (\tilde{z})$, inferred by someone who treats DM as dust, and, second, to express this function in terms of the actually-measured quantity, $z$. A~supporter of the collisionless-DM scenario, necessarily performs calculations in the framework of the EdS Universe, in which the luminosity distance is given by \be \tilde{d}_L (\tilde{z}) = \frac{2c}{\tilde{H}_0} \: \left ( 1 + \tilde{z} \right )^{1/2} \: \left [ \left ( 1 + \tilde{z} \right )^{1/2} - 1 \right ] \: . \label{eq40}\ee 

According to Kleidis and Spyrou \cite{159}, in the context of the iDMF approach, Equation~(\ref{eq40}) is translated to \be \tilde{d}_L (z) = \frac{2c}{\left ( 1 - 4w \Om_M \right ) H_0} \: (1 + z)^{\frac{1}{2} (1 - 3w\Om_M)} \left [ ( 1 + z)^{\frac{1}{2} (1 - 3w\Om_M)} - 1 \right ] \: . \label{eq41}\ee 

\begin{figure}[ht!]
\centerline{\mbox {\epsfxsize=15.cm \epsfysize=10.cm
\epsfbox{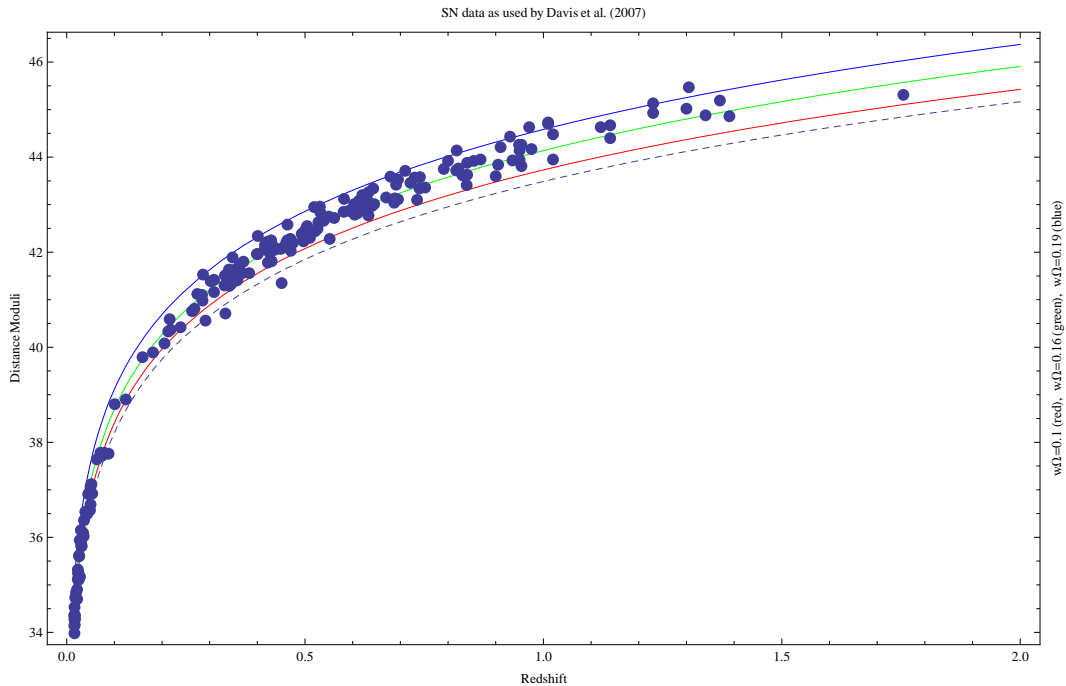}}} 
\caption{Overplotted on the Hubble diagram of the SNe Ia sample used by Davis \emph{et al}. \cite{181}, are the theoretical curves of the distance modulus in the iDMF model, $\mu (z)$, for $w \Om_M = 0.10$ (red solid line), $w \Om_M = 0.16$ (green line), and $w \Om_M = 0.19$ (blue solid line). The dashed line represents the theoretical curve associated to the distance modulus in the context of the collisionless-DM approach.}\label{fig1}
\end{figure}

However, in depicting Equation~(\ref{eq38}) on the $\mu$ \emph{versus} $z$ diagram of a sample of SNe events, an observer who treats DM as collisionless, unavoidably misinterprets the measured quantity $z$ as $\tilde{z}$ (also, the present-time value of the Hubble parameter, $H_0$, is misinterpreted as $\tilde{H}_0$). In other words, the theoretical formula of the luminosity distance that is used by someone who treats DM as dust, is (falsely) written in the form \be \tilde{d}_L (\tilde{z}) = \frac{2c}{H_0} \: \left ( 1 + z \right )^{1/2} \: \left [ \left ( 1 + z \right )^{1/2} - 1 \right ] \: , \ee instead of that given by the combination of Equations~(\ref{eq35}) and (\ref{eq41}). This is most prominently demonstrated in Figure \ref{fig1}, where, on the Hubble diagram of the SNe Ia dataset used by \mbox{Davis \emph{et~al}.~\cite{181},} we have overplotted the theoretically determined (in the context of the iDMF model) function $\mu (z)$, for three values of the combination $w \Om_M$, namely, $w \Om_M = 0.10$ (red solid line), $w \Om_M = 0.16$ (green solid line) and $w \Om_M = 0.19$ (blue solid line), together with the corresponding collissionless-DM quantity, $\tilde{\mu} (z)$ (dashed line). We observe that, when the thermodynamical content of the iDMF model is taken into account, the theoretically-determined distance modulus fits the entire SNe dataset under consideration quite accurately.

We see that, in the iDMF framework, provided that cosmologists no longer insist on adopting the collisionless-DM approach, the observed dimming of the SNe Ia standard candles would be only apparent. In other words, in the iDMF model there is no disagreement between observations and the theoretical prediction of the distant SNe Ia distribution.

\subsection{The Apparent Acceleration of the iDMF Model}

According to Kleidis and Spyrou \cite{159}, the Hubble parameter that is realized by a supporter of the collisionless-DM scenario, $\tilde{H}$, is given by \be \tilde{H} = H \: \frac{d}{dz} \left ( \frac{1+z}{f} \right ) = H_0 (1 + z)^{\frac{3}{2} (1 + w \Om_M)} \frac{1 - 2 w \Om_M + 3 w \Om_M \ln (1 + z)}{(1 + w \Om_M [1 + 3 \ln (1 + z)])^2} \: , \ee where we have used also Equations~(\ref{eq23}) and (\ref{eq30}). With the aid of 
Equation~(\ref{eq32}), $\tilde{H} (z)$ can be expressed in terms of $\tilde{z}$, as follows \be \tilde{H} = H_0 \: (1 + \tilde{z})^{\frac{3 (1 + w \Om_M)}{2 (1 - 3 w \Om_M)}} \: (1 - 3 w \Om_M) \frac{1 - 5 w \Om_M + 3 w \Om_M \ln (1 + \tilde{z}) + \textsl{O}(w \Om_M)^2}{ \left [ 1 - 2 w \Om_M + 3 w \Om_M \ln (1 + \tilde{z}) + \textsl{O}(w \Om_M)^2 \right ]^2} \: . \label{eq44}\ee 

In the context of the collisionless-DM approach the deceleration parameter, $\tilde{q}$, is defined as \be \tilde{q} (\tilde{z}) = \frac{d \tilde{H}/d \tilde{z}}{\tilde{H} (\tilde{z})} ( 1 + \tilde{z}) - 1, \ee which, upon consideration of Equation~(\ref{eq44}), yields \be \tilde{q} (\tilde{z}) = \frac{1}{2} \: \fdot \: \left [ \frac{1 - 4 w \Om_M + 6 w \Om_M \ln (1 + \tilde{z}) + \textsl{O}(w \Om_M)^2}{1 - 10 w \Om_M + 6 w \Om_M \ln (1 + \tilde{z}) + \textsl{O}(w \Om_M)^2} \right ] . \ee 

To terms linear in $w \Om_M$, the condition for accelerated expansion, $\tilde{q} (\tilde{z}) < 0$, results in \be 1 - 14 w \Om_M + 12 w \Om_M \ln (1 + \tilde{z}) < 0 \: , \label{eq47}\ee from which it is evident that, as far as a supporter of the collisionless-DM scenario is concerned, $\tilde{q} (\tilde{z}) < 0$ at cosmological redshifts \be \tilde{z} < \tilde{z}_{tr} = e^{\frac{14 w \Om_M - 1}{12 w \Om_M}} - 1. \label{eq48}\ee 

Equation~(\ref{eq48}) suggests that, if the Universe matter content is treated as an isothermal DM fluid~with \be w \Om_M > w_c \Om_M = \frac{1}{14} \approx 0.0714 \ee (\emph{i.e.}, $w > w_c \approx 0.238$), then, from the point of view of an observer who persists in treating DM as collisionless, there exists a transition value, $\tilde{z}_{tr}$, of the cosmological redshift, below which, such a cosmological model is accelerating.

In view of all the above, if the universal gravitational field is driven by an isothermal DM fluid with thermodynamical content, then, what is inferred as acceleration of the cosmic expansion could be only apparent, based on the misinterpretation of several cosmological parameters, by those observers who (although living in the iDMF model) simply insist on treating DM as dust.

The combination of Equations~(\ref{eq32}) and (\ref{eq48}) results in a non-linear algebraic equation, which involves the transition value, $z_{tr}$, of the truly measured cosmological redshift, $z$, that is, \be \left ( 1 + z_{tr} \right ) \: e^{0.25/3 w \Om_M} = 3.2114 \left ( 1 + z_{tr} \right )^{3 w \Om _M} \: . \label{eq50}\ee 

Admitting that $z_{tr} = 0.752 \pm 0.046$ \cite{30}, Equation~(\ref{eq50}) can be solved numerically with respect to the combination $w \Om_M$, yielding 
\vspace{-1pt}
\be \left ( w \Om_M \right )_{tr} = 0.1062 \pm 0.0028 \: . \label{eq51}\ee 

In view of Equation~(\ref{eq51}), $w \geq \frac {1}{3}$, \emph{i.e.}, compatibility of the iDMF approach with the observational data currently available, suggests that the DM consists of relativistic particles. For the time being, a pure HDM model can not reproduce the large-scale structure of the Universe \cite{168}, although there are several scientists disagreeing with such a premise \cite{182,183}. We should also mention that, the alternative DE model so considered, reproduces to high accuracy (e.g., much more accurately than the $\Lm$CDM model itself) the observational results concerning the statistically-independent distance constraint associated with BAOs data \cite{101} (for a detailed analysis, see \cite{159}). 

The idea that the DE could be represented by the energy of the internal motions of an isothermal DM fluid is quite challenging and should be further examined in the search for alternatives to the DE concept. In this framework, and in an effort to confront with the HDM issue, in the next Section, we consider a cosmological model with matter content also in the form of a thermodynamically-involved DM fluid, the volume elements of which, this time, perform polytropic flows (pDMF model) \cite{160}. 

\section{Polytropic Processes in a Cosmological DM Fluid}\label{sec4}

In realistic astrophysics, polytropic processes are much more physically relevant than isothermal flows (see, e.g., \cite{184} (pp. 64--69)). On galactic scale, polytropic processes have been proved very useful in modeling dark galactic haloes, leading to a significant improvement of the galaxies velocity dispersion profiles \cite{161,162,163,164,165,166,167}. In the cosmological framework, polytropic (DM) models were first encountered as natural candidates for Cardassian cosmology (see, e.g., \cite{59,60,61}). They have been used also in the unified DE models, to describe a potential interaction between DM and DE (see, e.g.,~\cite{185,186,187,188,189,190,191,192,193,194,195}).

The polytropic process is a general way of treating flow motions, including many thermodynamic processes in a single formula. It is a reversible process, such that, the specific heat of a thermodynamical system, \be {\cal C} = \frac{dQ}{dT} \: , \ee varies in a prescribed manner (see, e.g., \cite{196} (p.~2)). In the special case where ${\cal C} = constant$, the thermodynamical system is left with only one independent state variable, the rest-mass density (barotropic flow). Accordingly, the fundamental equation of state, \be p \propto \rh T \: , \ee is decomposed to \be p = p_0 \left ( \frac{\rh}{ \rh_0} \right )^{\Gm} \label{eq54}\ee and \be T = T_0 \left ( \frac{\rh}{\rh_0} \right )^{\Gm - 1} \ee  (see, e.g., \cite{196} (p. 9), \cite{197} (p. 85)), where $p_0$ and $T_0$ denote the present-time values of pressure and temperature, respectively, and $\Gm$ is the polytropic exponent, defined as \be \Gm = \frac{{\cal C}_P - {\cal C}}{{\cal C}_V - {\cal C}} \ee (see, e.g., \cite{196} (p. 5), \cite{198} (p. 86)), where ${\cal C}_P$ (${\cal C}_V$) is the specific heat at constant pressure (volume). At this point, we need to stress that, for the definition of specific heats, the concept of equilibrium is essential. However, in an expanding Universe such a concept cannot be posed in an unambiguous way; hence, in a cosmological setup, the definition of specific heats may not coincide with the corresponding thermodynamic one (in connection, see \cite{198,199,200,201}). 

In the pDMF model, the first law of Thermodynamics yields \be {\cal U} = {\cal U}_0 \left ( \frac{\rh}{ \rh_0} \right )^{\Gm - 1} , \label{eq57}\ee where \be {\cal U}_0 = {\cal C} T_0 + \frac{1}{\Gm - 1} \frac{p_0}{\rh_0} \ee is the present-time value of the cosmic fluid internal energy. In this case, the continuity Equation~(\ref{eq14}) results in \be \Gm {\cal U}_0 \left ( \dot{\rh} + 3 \frac{\dot{S}}{S} \rh \right ) + \dot{\cal E} + 3 \frac{\dot{S}}{S} {\cal E} - 3 (\Gm - 1) \rh_0 {\cal C} T_0 \frac{\dot{S}}{S} \left ( \frac{\rh}{\rh_0} \right )^{\Gm} = 0 \: , \label{eq59}\ee where the dot denotes differentiation with respect to cosmic time, $t = \int^{\et} S(\et) d \et$. Recall that, by definition, each pDMF volume element is assumed to be a closed system, \emph{i.e.}, the total number of its particles is conserved, so that \be \dot{\rh} + 3 \frac{\dot{S}}{S} \rh = 0 \: . \label{eq60}\ee 

According to Equation~(\ref{eq60}), the evolution of the rest-mass density in the pDMF model is (once again) given by Equation~(\ref{eq14}). Now, Equation~(\ref{eq59}) results in \be \dot{\cal E} + 3 \frac{\dot{S}}{S} {\cal E} - 3 (\Gm - 1) \rh_0 {\cal C} T_0 \frac{\dot{S}}{S} \left ( \frac{S_0}{S} \right )^{3 \Gm} = 0 \: , \ee yielding \be {\cal E} = \rh_0 c^2 \left ( \frac{S_0}{S} \right )^3 - \rh_0 {\cal C} T_0 \left ( \frac{S_0}{S} \right )^{3 \Gm} . \label{eq62}\ee 

In view of Equations~(\ref{eq14}), (\ref{eq57}) and (\ref{eq62}), the total energy density of the pDMF model is written in the form \be \varep = \rh_0 c^2 \left ( \frac{S_0}{S} \right )^3 + \frac{p_0}{\Gm - 1} \left ( \frac{S_0}{S} \right )^{3 \Gm} = \rh c^2 + \frac{p}{\Gm -1} \: , \label{eq63}\ee upon consideration of which, the evolution of a spatially-flat pDMF model is determined by the solution of the Friedmann equation \be \left ( \frac{H}{H_0} \right )^2 = \Om_M \left ( \frac{S_0}{S} \right )^3 \left [ 1 + \frac{1}{\Gm - 1} \frac{p_0}{\rh_0 c^2} \left ( \frac{S_0}{S} \right )^{3 (\Gm - 1)} \right ] . \label{eq64}\ee 

Extrapolation of Equation~(\ref{eq64}) to the present epoch, yields the present-time value of the isotropic pressure, as \be p_0 = \rh_0 c^2 (\Gm - 1) \frac{1 - \Om_M}{\Om_M} \: . \label{eq65}\ee 

For $\Gm < 1$, Equation~(\ref{eq65}) suggests that, the pressure of a polytropic-DM perfect fluid is negative. In this case, the quantity $\varep + 3 p$ may also become negative, leading to $\ddot{S} > 0$ (see, e.g., \cite{202}). In other words, the pDMF model with $\Gm < 1$ may accelerate its expansion.

Upon consideration of Equation~(\ref{eq65}), Equation~(\ref{eq64}) is written in the form \be \left ( \frac{H}{H_0} \right )^2 = \left ( \frac{S_0}{S} \right )^3 \left [ \Om_M + (1 - \Om_M) \left ( \frac{S}{S_0} \right )^{3 (1 - \Gm)} \right ] \label{eq66}\ee and Equation~(\ref{eq63}) results in 
\vspace{-6pt}
\be \varep = \rh_c c^2 \left [ \Om_M \left ( \frac{S_0}{S} \right )^3 + (1 - \Om_M) \left ( \frac{S_0}{S} \right )^{3 \Gm} \right ] \: . \label{eq67}\ee 

As a consequence, in the pDMF model, the present-time value of the total energy density parameter equals to unity, \emph{i.e.}, \be \Om_0 =  \frac{\varep_0}{\varep_c} = \frac{\rh_c c^2}{\rh_c c^2} \left [ \Om_M + (1 - \Om_M) \right ] = 1 \: . \label{eq68}\ee 

In view of Equations~(\ref{eq65}) and (\ref{eq68}), the pDMF Universe with $\Gm < 1$ might be an elegant solution to the DE problem, by addressing both spatial flatness and the accelerated expansion in a unique theoretical framework. For this reason, in what follows, we shall scrutinize the pDMF model \mbox{with $\Gm < 1$.}

By virtue of Equation~(\ref{eq67}), the rest-mass energy density, $\varep_{mat} = \rh c^2$, and the internal (dark) energy density, $\varep_{int} = \varep - \varep_{mat}$, of the Universe matter-energy content satisfy the relation \be \frac{\varep_{int}}{\varep_{mat}} = \frac{1 - \Om_M}{\Om_M} \frac{1}{(1+z)^{3(1 - \Gm)}} \: , \label{eq69}\ee which, at the present epoch $(z = 0)$, results in \be \left . \frac{\varep_{int}}{\varep_{mat}} \right \vert_0 = \frac{1 - \Om_M}{\Om_M} \: .\label{eq70} \ee 

On the other hand, for $\Om_M = 0.274$ \cite{40}, Equation~(\ref{eq65}) suggests that, today, \mbox{$p_0 = - 2.650 (1 - \Gm) \rh_0 c^2$.} This result might lead to the assumption that the pDMF model is, in fact, a phantom Universe, where $p_0 < - \varep_0$. Today, several observational data indicate that the basic cosmic ingredient might (very well) consist of phantom DE (see, e.g., \cite{41,203}). However, the latest Planck results suggest that, this is probably due to a geometric degeneracy, which will be erased as more data are added~\cite{204}. In~view of such a perspective, we note that, in the pDMF model, the total energy density at the present epoch is not given by $\rh_0 c^2$, but by $\varep_0 = \Om_M^{-1} \rh_0 c^2$ (\emph{cf.} 
Equation~(\ref{eq67})). Accordingly, Equation~(\ref{eq65}) results~in \be p_0 = - (1 - \Gm) (1 - \Om_M) \varep_0 \: , \ee from which, we deduce that $p_0 > - \varep_0$, as long as \be \Gm > - \frac{\Om_M}{1 - \Om_M} \cong - 0.377 \: . \label{eq72}\ee 

Clearly, the pDMF model with $-0.377 < \Gm < 1$ does not fall into the realm of phantom~cosmology. 

\subsection{Aleviating the Age Problem of the Universe}

In the pDMF model under consideration, Equation~(\ref{eq66}) reads \be \left [ \frac{d}{d t} \left ( \frac{S}{S_0} \right )^{3/2} \right ]^2 = \frac{1}{t_{EdS}^2} \left \lbrace \Om_M + (1 - \Om_M) \left [ \left ( \frac{S}{S_0} \right )^{3/2} \right ]^{2 (1 - \Gm)} \right \rbrace \: , \label{eq73}\ee where $t_{EdS} = \frac{2}{3 H_0}$ is the age of the Universe in the EdS model. Equation~(\ref{eq73}), can be solved explicitly in terms of hypergeometric functions, $_2F_1 (a \: , \: b \: ; \: c \: ; \: x)$, of a complex variable, $x$ (see, e.g., \cite{205} (pp.~1005--1008)), as follows 
\vspace{-6pt}
\be \left ( \frac{S}{S_0} \right )^{\frac{3}{2}} \: _2F_1 \left ( \frac{1}{2 (1 - \Gm)} \: , \: \frac{1}{2} \: ; \: \frac{3 - 2 \Gm}{2 (1 - \Gm)} \: ; - \left ( \frac{1 - \Om_M}{\Om_M} \right ) \left [ \frac{S}{S_0} \right ]^{3 (1 - \Gm)} \right ) = \sqrt{\Om_M} \left ( \frac{t}{t_{EdS}} \right ) \: . \label{eq74}\ee 

Since $a + b = \frac{1}{2 (1 - \Gm)} + \frac{1}{2} < \frac{3 - 2 \Gm}{2 (1 - \Gm)} = c$, the hypergeometric series involved, converges absolutely within the unit circle $\left \vert \frac{S}{S_0} \right \vert \leq 1$, for every value of $\Gm < 1$ (see, e.g., \cite{206} (p. 556)). For $\Om_M = 1$, Equation~(\ref{eq74}) yields $S = S_0 \left ( \frac{t}{t_{EdS}} \right )^{2/3}$, \emph{i.e.}, the scale factor of the EdS model, as it should. On the other hand, in the (isobaric) $\Gm = 0$ case, Equation~(\ref{eq74}) is reduced to \be \left ( \frac{S}{S_0} \right )^{\frac{3}{2}} \: _2F_1 \left ( \frac{1}{2} \: , \: \frac{1}{2} \: ; \: \frac{3}{2} \: ; \: - \left ( \frac{1 - \Om_M}{\Om_M} \right ) \left [ \frac{S}{S_0} \right ]^3  \right ) = \sqrt{\Om_M} \left ( \frac{t}{t_{EdS}} \right ) \: , \ee which, upon consideration of the identity \be _2F_1 \left ( \frac{1}{2} \: , \: \frac{1}{2} \: ; \: \frac{3}{2} \: ; \: - x^2  \right ) = \frac{1}{x} \sinh^{-1} (x) \ee (\emph{cf.} \cite{205}, Equation 9.121.28, (p. 1007), \cite{206}, Equation 15.1.7, (p. 556)), results in \be S(t) = S_0 \left ( \frac{\Om_M}{1 - \Om_M} \right )^{1/3} \sinh^{2/3} \left ( \sqrt{1 - \Om_M} \frac{t}{t_{EdS}} \right ) \: . \label{eq77}\ee    

For $1 - \Om_M = \Om_{\Lm}$, Equation~(\ref{eq77}) represents the scale factor of the $\Lm$CDM model (\emph{cf.} Equation~(5) of \cite{207}), as it (once again) should. 

\begin{figure}[ht!]
\centerline{\mbox {\epsfxsize=15.cm \epsfysize=10.cm
\epsfbox{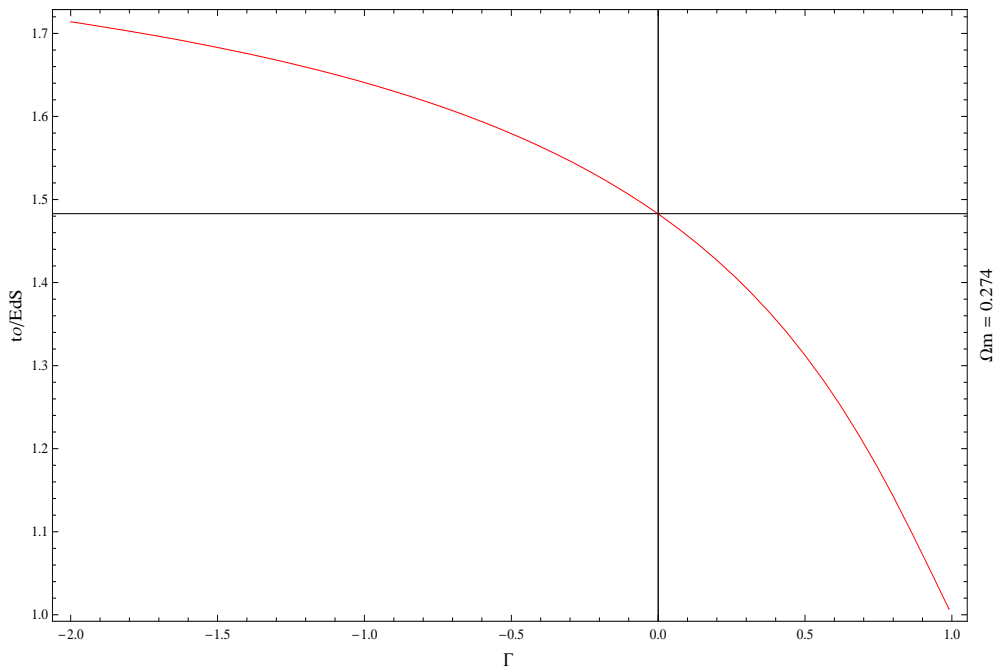}}} 
\caption{The age of the DM fluid perform polytropic flows (pDMF) model, $t_0$, in units of $t_{EdS}$, as a function of the polytropic exponent $\Gm < 1$ (red solid line). For every $\Gm < 1$, $t_0 > t_{EdS}$, and $t_0$ approaches $t_{EdS}$ only in the isothermal $(\Gm \rarrow 1)$ limit. The horizontal solid line denotes the age of the Universe $(t_0 = 1.483 \; t_{EdS})$ in the isobaric $(\Gm = 0)$ $\Lm$CDM limit of the pDMF model.}\label{fig2}
\end{figure}

Using Equation~(\ref{eq74}), we can also determine the age of the Universe, $t_0$, in the context of the pDMF approach. In units of $t_{EdS}$, it is given by \be \frac{t_0}{t_{EdS}} = \frac{1}{\sqrt {\Om_M}} \: _2F_1 \left ( \frac{1}{2 (1 - \Gm)} \: , \: \frac{1}{2} \: ; \: 1 + \frac{1}{2 (1 - \Gm)} \: ; \: - \frac{1 - \Om_M}{\Om_M} \right ) \: ,\label{eq78} \ee the behaviour of which, as a function of the polytropic exponent $\Gm < 1$, is presented in Figure \ref{fig2}. In~the isobaric $(\Gm = 0)$ case, Equation~(\ref{eq78}) yields \be t_0 = t_{EdS} \frac{1}{\sqrt{1 - \Om_M}} \sinh^{-1} \sqrt{\frac{1 - \Om_M}{\Om_M}} \: , \ee which, for $\Om_M = 0.274$ \cite{40}, results in $t_0 = 1.483 \; t_{EdS} = 13.778 \; Gys$. This value coincides to the corresponding {\em nine-year WMAP} result \cite{41} and lies well-within range of its (latest) {\it Planck} counterpart~\cite{204}, concerning the age of the $\Lm$CDM Universe. Clearly, the pDMF model does not suffer from the (so-called) age problem. 

\begin{figure}[ht!]
\centerline{\mbox {\epsfxsize=15.cm \epsfysize=10.cm
\epsfbox{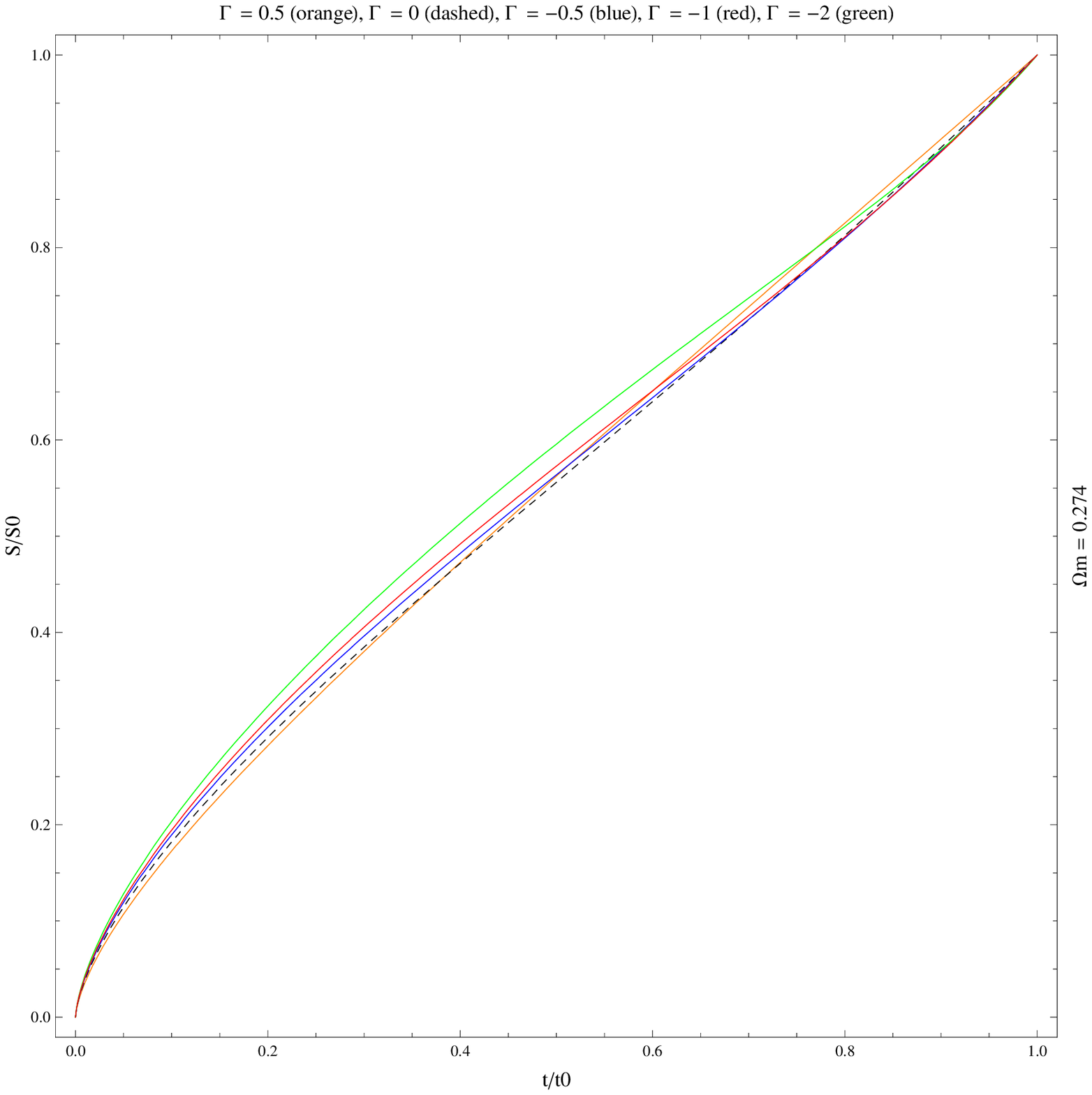}}} 
\caption{The scale factor, $S$, of the pDMF model with $\Om_M = 0.274$ (in units of its present-time \mbox{value, $S_0$),} as a function of the cosmic time $t$ (in units of $t_0$), for $\Gm = 0.5$ (orange), $\Gm = 0$ (dashed), $\Gm = -0.5$ (blue), $\Gm = - 1$ (red), and $\Gm = - 2$ (green). For each and every curve, there is a value of $t < t_0$, above which $S(t)$ becomes concave, \emph{i.e.}, the Universe accelerates its expansion.} \label{fig3}
\end{figure}

Eventually, from the combination of Equations~(\ref{eq74}) and (\ref{eq78}), we find that, the scale factor of the pDMF model is given by \be \left ( \frac{S}{S_0} \right )^{3/2} \frac{_2F_1 \left ( \frac{1}{2 (1 - \Gm)} \: , \: \frac{1}{2} \: ; \: \frac{3 - 2 \Gm}{2 (1 - \Gm)} \: ; \: - \left ( \frac{1 - \Om_M}{\Om_M} \right ) \left [ \frac{S}{S_0} \right ]^{3 (1 - \Gm)} \right )}{_2F_1 \left ( \frac{1}{2 (1 - \Gm)} \: , \: \frac{1}{2} \: ; \: \frac{3 - 2 \Gm}{2 (1 - \Gm)} \: ; \: - \frac{1 - \Om_M}{\Om_M} \right )} = \frac{t}{t_0} \: , \ee the time behaviour of which, for several values of $\Gm < 1$, is presented in Figure \ref{fig3}. We note that, there is always a value of $t < t_0$, above which, the function $S(t)$ becomes concave, \emph{i.e.}, $\ddot{S} > 0$. This is a very important result, suggesting that, the pDMF model with $\Gm < 1$ accelerates its expansion. This can be readily confirmed, upon the calculation of the deceleration parameter associated to this model.

\subsection{Confronting with the Coincidence Problem}

In the pDMF model with $\Gm < 1$, the Hubble parameter Equation~(\ref{eq66}), in terms of the cosmological redshift, is written in the form \be H = H_0 ( 1 + z )^{\frac{3}{2}} \left [ \Om_M + \frac{1 - \Om_M}{( 1 + z )^{3 (1 - \Gm)}} \right ]^{1/2} . \label{eq81}\ee 

Accordingly, the corresponding deceleration parameter Equation~(\ref{eq24}), yields \be q (z) = \frac{1}{2} \left [ 1 - \frac{3 (1 - \Gm) (1 - \Om_M)}{\Om_M (1 + z)^{3 (1 - \Gm)} + (1 - \Om_M)} \right ] \: .\label{eq82} \ee 

For $z \gg 1$ (\emph{i.e.}, in the distant past), $q \rarrow \frac{1}{2}$ and the Universe behaves as the EdS model, \emph{i.e.}, a dust (in other words, decelerating) FRW model. On the other hand, for $z = 0$ (\emph{i.e.}, at the present epoch), we have \be q_0 = \frac{1}{2} \left [ 1 - 3 (1 - \Gm) (1 - \Om_M) \right ] \: . \label{eq83}\ee 

The minus sign on the rhs of Equation~(\ref{eq82}) suggests that there is a transition value of $z$, namely, $z_{tr}$, below which, $q(z)$ becomes negative, \emph{i.e.}, the Universe accelerates its expansion. It is given by \be z_{tr} = \left [ (2 - 3 \Gm) \frac{1 - \Om_M}{\Om_M} \right ]^{\frac{1}{3 (1 - \Gm)}} - 1 \: . \label{eq84}\ee 

For $\Gm = 0$, Equation~(\ref{eq84}) yields $z_{tr} = 0.744$, which lies well-within range of the corresponding $\Lm$CDM result, namely, $z_{tr} = 0.752 \pm 0.041$ \cite{30}. In view of Equation~(\ref{eq84}), the condition $z_{tr} \geq 0$ imposes the following constraint on the potential values of $\Gm$, \be \Gm \leq \frac{1}{3} \left [ 2 - \frac{\Om_M}{1 - \Om_M} \right ] \: . \label{eq85}\ee 

\begin{figure}[ht!]
\centerline{\mbox {\epsfxsize=15.cm \epsfysize=10.cm
\epsfbox{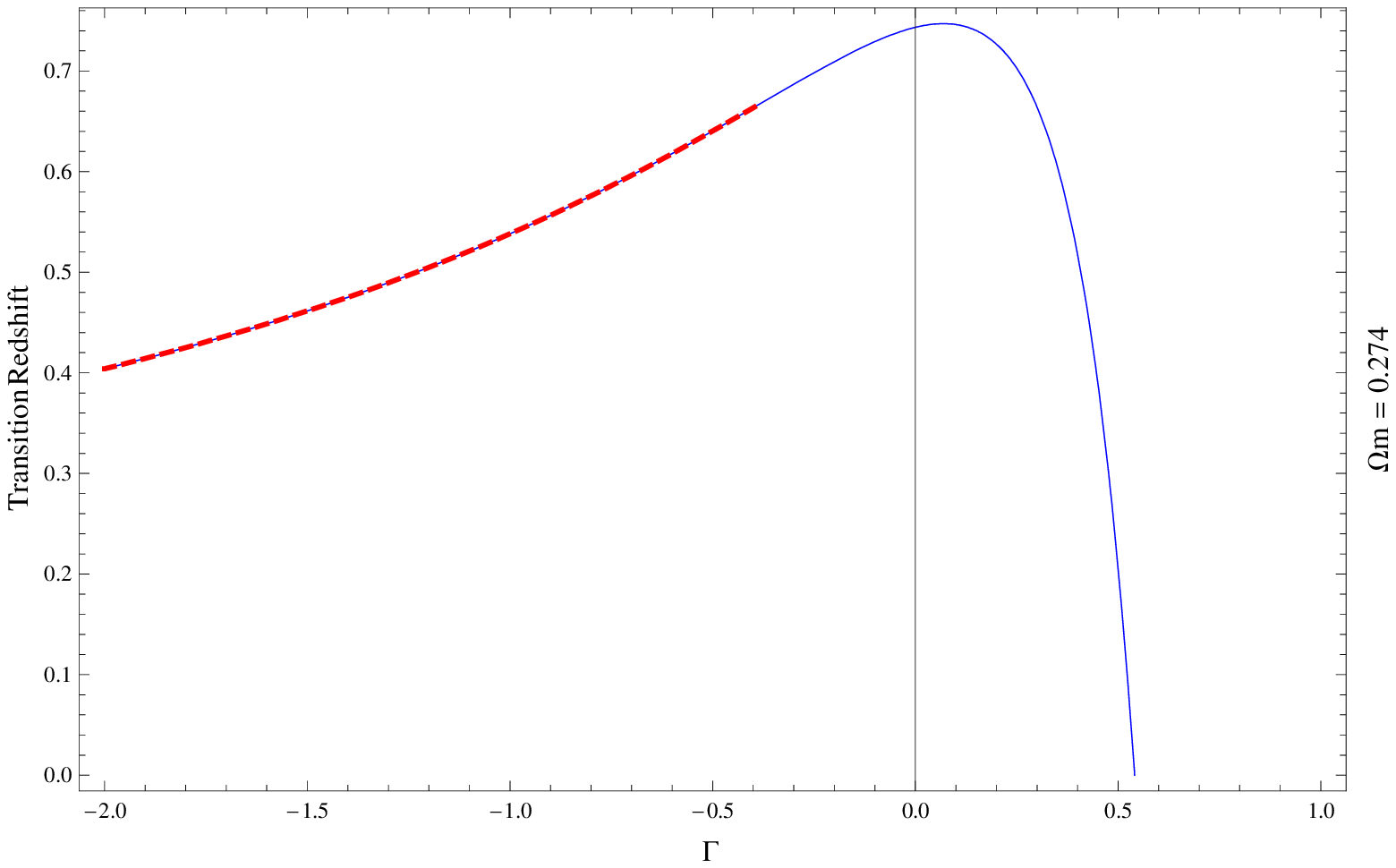}}} 
\caption{The transition redshift, $z_{tr}$, in the pDMF model as a function of the polytropic exponent, $\Gm$ (blue solid curve). For $\Gm < -0.377$, the Universe resides to the phantom realm (red dashed curve).} \label{fig4}
\end{figure}

For $\Om_M = 0.274$ \cite{40}, Equation~(\ref{eq85}) yields the upper limit, $\Gm \leq 0.540$, while the requirement for a non-phantom Universe, $\Gm > -0.377$, may serve as a lower bound of $\Gm$. The behaviour of $z_{tr}$, as a function of $\Gm \leq 0.540$, is presented in Figure \ref{fig4}.

We see that, polytropic acceleration is definitely not a coincidence. The pDMF model with $- 0.377 < \Gm \leq 0.540$, most naturally accelerates its expansion, at cosmological redshifts lower than the transition value given by Equation~(\ref{eq84}), without the need for either any exotic DE or the cosmological constant. The question that arises now is, whether these theoretical results are confirmed also by the observational data related to the distant SNe Ia standard candles or not.

\subsection{Compatibility with the Recent SNe Ia Data}

Today, many samples of SNe Ia data are used, to scrutinize the viability of the various DE models (see, e.g., \cite{181}). The most extended one is the Union 2.1 Compilation \cite{30}, which consists of 580 SNe Ia events (Available at http://www.supernova.lbl.gov/Union). Accordingly, to estimate the compatibilty of the pDMF model with the observational data associated to the SNe Ia distant indicators, once again, we overplot the corresponding theoretically-derived distance modulus on the Hubble diagram of the Union 2.1 SN Compilation. In this case, the luminosity distance is given by \be d_L (z) = c (1+z) \int_0^z \frac{d z^{\prime}}{H(z^{\prime})} \label{eq86}\ee (see, e.g., \cite{170} (p. 76)), where $H(z)$ is given by Equation~(\ref{eq81}). Equation~(\ref{eq86}) can be solved explicitly in terms of hypergeometric functions (see, e.g., \cite{205} 
(pp. 1005--1008)), resulting in \bea && d_L (z) = \frac{2 c}{H_0} \frac{1}{\sqrt {1 - \Om_M}} \frac{1 + z}{2 - 3 \Gm} \left [ (1 + z)^{\frac{2 - 3 \Gm}{2}} \times \right . \nn \\ && \left . _2F_1 \left ( \frac{2 - 3 \Gm}{6 ( 1 - \Gm)}  \: , \: \frac{1}{2} \: ; \: \frac{8 - 9 \Gm}{6 ( 1 - \Gm)} \: ; \: - \left [ \frac{\Om_M}{1 - \Om_M} \right ] (1 + z)^{3 (1 - \Gm)} \right )  - \right . \nn \\ && \left . _2F_1 \left ( \frac{2 - 3 \Gm}{6 (1 - \Gm)}  \: , \: \frac{1}{2} \: ; \: \frac{8 - 9 \Gm}{6 ( 1 - \Gm)} \: ; \: - \left [ \frac{\Om_M}{1 - \Om_M} \right ] \right ) \right ] \: . \label{eq87}\eea 

\begin{figure}[ht!]
\centerline{\mbox {\epsfxsize=15.cm \epsfysize=10.cm
\epsfbox{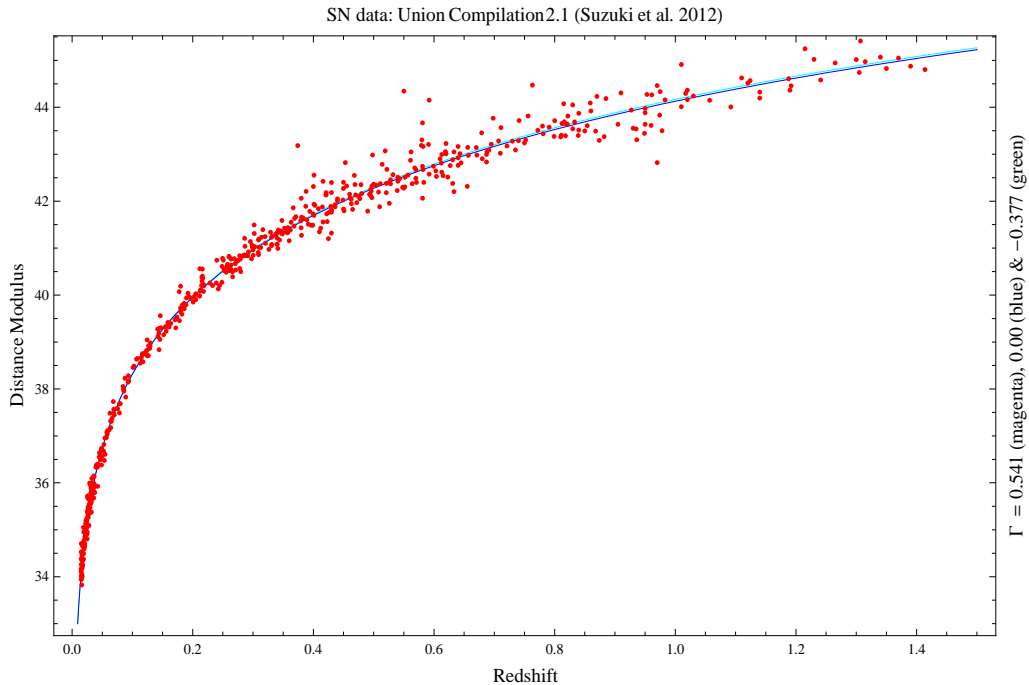}}} 
\caption{Overplotted to the Hubble diagram of the Union 2.1 Compilation are the best-fit curves (too close to be distinguished) representing the function $\mu (z)$ in the pDMF model, for $-0.09 < \Gm \leq 0$.}\label{fig5}
\end{figure}

By virtue of Equation~(\ref{eq87}), the function $\mu (z)$, given by Equation~(\ref{eq37}), is overplotted on the $\mu$ \emph{versus} $z$ diagram of the extended Union 2.1 Compilation \cite{30}. The outcome is depicted in Figure \ref{fig5}, for $-0.09 < \Gm \leq 0$ (in connection, see Section 4.4). It is evident that, the theoretical curve representing the distance modulus in the context of the pDMF model, fits the entire dataset quite accurately. In other words, there is no disagreement between our theoretical prediction and the observed distribution of the distant SNe Ia events. 

\subsection{Determining the Value of the Polytropic Exponent}

In relativistic hydrodynamics, the isentropic velocity of sound is defined as \be c_s^2 = c^2 \left ( \frac{\partial p}{\partial \varep} \right )_{\cal S} \label{eq88}\ee (see, e.g., \cite{208} (p. 52)), where $\left ( \frac{\partial p}{\partial \varep} \right )_{\cal S} \leq 1$, in order to avoid violation of causality \cite{209}. In view of Equation~(\ref{eq88}), the barotropic flow in the pDMF model yields a velocity of sound that is not constant, but (rather) a function of the cosmological redshift, parametrized by $\Gm$. A velocity of sound that varies in a prescribed manner, could help us restore the degeneracy between the constituents of the dark sector in the unified DE models (see, e.g., \cite{201,210,211,212}). It may also reveal the functional form of the coupling parameter in the interactive DE models (see, e.g., \cite{213,214,215}). In fact, as regards the former class of models, a constant velocity of sound would require $c_s^2 \rarrow 0$, in order to match the (baryon) mass power spectrum to the SDSS DR7 data \cite{158}. Accordingly, we distinguish two cases:

\begin{enumerate}

\item[(i)] $\Gm = 0$: In this case, $p = constant = p_0$, and, therefore, \be c_s^2 (\Gm = 0) = 0 \: . \label{eq89}\ee In other words, in the isobaric $(\Gm = 0)$ limit, the pDMF model does resemble the $\Lm$CDM model, in which the cosmological constant does not carry any perturbations. For this reason, the $\Gm = 0$ case is often referred to as the $\Lm$CDM limit of the pDMF model.

\item[(ii)] $\Gm \neq 0$: In this case, the total energy density of the Universe matter-energy content (\ref{eq63}) is written in the form \be \varep = \underbrace{\rh c^2}_{\varep_{mat}} + \underbrace{\frac{p}{\Gm - 1}}_{\varep_{int}} = \rh_0 c^2 \left (\frac{p}{p_0} \right )^{1/\Gm} + \frac{p}{\Gm - 1} \: , \ee the partial differentiation of which, with respect to $\varep$, yields \be \left ( \frac{\partial p}{\partial \varep} \right )_{\cal S} = \frac{\Gm \left ( \frac{p}{\rh c^2} \right )}{1 + \frac{\Gm}{\Gm - 1} \left ( \frac{p}{\rh c^2} \right )} = \left ( \frac{c_s}{c} \right )^2 \: . \ee Accordingly, the velocity of sound in the pDMF model, as a function of the cosmological redshift, is given by \be \left ( \frac{c_s}{c} \right )^2 =  - \frac{\Gm (1 - \Gm) \frac{1 - \Om_M}{\Om_M}}{(1 + z)^{3(1 - \Gm)} + \Gm \frac{1 - \Om_M}{\Om_M}} \: , \ee in view of which, the requirement for a positive velocity-of-sound square yields a major constraint on the upper bound of $\Gm$, namely, \be \left ( \frac{c_s}{c} \right )^2 > 0 \Leftrightarrow \Gm < 0 \: . \label{eq93}\ee  

There are two values of $\left ( \frac{c_s}{c} \right )^2$ of particular interest, namely, (a) at transition $(z = z_{tr})$, where \be \left ( \frac{c_s}{c} \right )_{tr}^2 = \frac{\vert \Gm \vert}{2} \Rightarrow \vert \Gm \vert = 2 \left ( \frac{c_s}{c} \right )_{tr}^2 \: , \ee attributing to the polytropic exponent an unexpected physical interpetation, and \texttt{(b)} at the present epoch $(z = 0)$, when $\left ( \frac{c_s}{c} \right )^2$ attains its maximum value, namely, 
\vspace{-12pt}
\be \left ( \frac{c_s}{c} \right )_0^2 = \left ( 1 + \vert \Gm \vert \right ) \frac{\vert \Gm \vert \frac{1 - \Om_M}{\Om_M}}{1 - \vert \Gm \vert \frac{1 - \Om_M}{\Om_M}} \: . \ee 

Notice that, for a DM fluid consisting of relativistic particles (\emph{i.e.}, HDM), the velocity of sound would be $\left ( \frac{c_s}{c} \right )^2 = \frac{1}{3}$ (see, e.g., \cite{208} (p. 51), \cite{216} (p. 509)). Accordingly, the requirement for CDM at the present epoch constrain us to impose \be \left ( \frac{c_s}{c} \right )_0^2 < \frac{1}{3} \: , \label{eq96}\ee which, in the pDMF model under consideration, results in \be \vert \Gm \vert < \frac{2}{3} \left [ \sqrt{1 + \frac{3}{4} \frac{\Om_M}{1 - \Om_M}} - 1 \right ] \: = 0.089 \: . \label{eq97}\ee We see that, the physical requirements given by Equations~(\ref{eq93}) and (\ref{eq96}), together with Equation~(\ref{eq89}), have resulted in a narrower range of values of the polytropic exponent that can be attributed to a realistic pDMF model, namely, \be - 0.09 < \Gm \leq 0 \: . \label{eq98}\ee 

In view of Equation~(\ref{eq98}), in a pDMF model compatible to modern observational cosmology, the polytropic exponent, if not zero~(\emph{i.e.}, the $\Lm$CDM model), is definitely negative and very close to zero. 
Equation~(\ref{eq98}) is in good agreement to the corresponding result that arises for a generalized Chaplygin gas, $p \sim - \rh^{- \al}$, from the combination of X-ray and SNe Ia measurements with data from Fanaroff-Riley type IIb radio-galaxies, namely, \mbox{$\al = -0.09_{-0.33}^{+0.54}$ \cite{217}.} In addition, by virtue of Equation~(\ref{eq98}), the present-time value of the deceleration parameter given by Equation~(\ref{eq83}) falls into the range $-0.686 < q_0 \leq -0.589$, which lies at the lower part of the observationally-determined range of values for $q_0$, based on the SALT2 fitting to the SNe+BAO/CMB data, \emph{i.e.}, \mbox{$q_0 = -0.53_{-0.13}^{+0.17}$ \cite{218}.} 

\end{enumerate}

\subsection{Transition of the pDMF Model to Acceleration}

For $\Gm < 0$, the combination of Equations~(\ref{eq69}) and (\ref{eq84}) yields \be \frac{\varep_{int}}{\varep_{mat}} = \frac{1}{2 + 3 \vert \Gm \vert} \left ( \frac{1 + z_{tr}}{1 + z} \right )^{3(1 + \vert \Gm \vert)} , \ee which, at $z = z_{tr}$, results in \be \left. \frac{\varep_{int}}{\varep_{mat }} \right \vert_{tr} = \frac{1}{2 + 3 \vert \Gm \vert} \: . \label{eq100}\ee 

Equation~(\ref{eq100}) suggests that, in contrast to the common perception, the onset of transition from deceleration to acceleration in the pDMF model does not necessarily requires $\varep_{int} > \varep_{mat}$. In~fact, according to Equation~(\ref{eq69}), the internal (dark) energy density became equal to its rest-mass counterpart quite later, at $z = 0.384$ (in a model with $\Gm = 0$), which is remarkably close to the observationally-determined value $z = 0.391 \pm 0.033$ \cite{30}, associated with the $\Lm$CDM model. According to Equation~(\ref{eq100}), for values of the polytropic exponent in the range given by Equation~(\ref{eq98}), the transition from deceleration to acceleration took place when \be 0.44 < \frac{\varep_{int }}{\varep_{mat}} \leq 0.50 \: . \ee 

The question is, why does it happen in this way. The answer is both revealing and simple: Because of the GR itself! 

In the context of GR, the dynamics of a spatially-flat FRW model is most appropriately determined by 
Equations~(\ref{eq7}) and (\ref{eq10}). In terms of cosmic time, the combination of these two equations~yields \be \frac{\ddot{S}}{S} = - \frac{4 \pi G}{3 c^2} \left ( \eps + 3 p \right ) \ee (see, e.g., \cite{68,202}); hence, the condition for accelerated expansion, $\ddot{S} > 0$, results in \be \eps + 3 p < 0 \: . \label{eq103}\ee 

In terms of the pDMF approach (\emph{cf.} Equations~(\ref{eq54}) and (\ref{eq63})), the condition Equation~(\ref{eq103}) is written in the form \be \varep + 3 p =  \rh_0 c^2 (1 + z)^3 \left [ 1 - (2 + 3 \vert \Gm \vert) \frac{1 - \Om_M}{\Om_M} \frac{1}{(1 + z)^{3(1+ \vert \Gm \vert)}} \right ] < 0 \: . \label{eq104}\ee 

According to Equation~(\ref{eq104}), the pDMF model under consideration accelerates its expansion at cosmological redshifts lower than a specific value, namely, \be z < \left [ (2 + 3 \vert \Gm \vert) \frac{1 - \Om_M}{\Om_M} \right ]^{\frac{1}{3(1+ \vert \Gm \vert)}} - 1 \equiv z_{tr} \: , \label{eq105}\ee in complete agreement to the transition redshift, given by Equation~(\ref{eq84}). In view of Equations~(\ref{eq104}) and (\ref{eq105}), the pDMF model could most definitely explain why the Universe transits to acceleration at $z_{tr}$, without the need for an extra (dark) energy component or the cosmological constant. Instead, it would reveal a (conventional) form of DE that so far has been disregarded, \emph{i.e.}, due to the (polytropic) DM fluid internal motions. It is worth mentioning that the $\Lm$CDM limit $(\Gm = 0)$ of the pDMF model reproduces to high accuracy (also) the value of the (so-called) shift parameter \cite{219}, obtained by fitting the CMB data to the standard $\Lm$CDM model \cite{220} (for a detailed analysis, see \cite{160}).

\section{Discussion and Conclusions}\label{sec5}

In this article, we review a series of recent theoretical results regarding a conventional approach to the DE concept. In particular, we have explored the possibility that, the DE needed to flatten the Universe and to accelerate its expansion, is attributed to the energy of the cosmic fluid internal motions. In this framework, the Universe is filled with a perfect fluid, consisting mainly of self-interacting dark matter, the volume elements of which perform hydrodynamic flows. The~pressure of this fluid is given by a barotropic equation of state, the functional form of which depends on the type of thermodynamic processes occuring in its interior. Accordingly, we have distinguished two cases \cite{159,160}.

In the first case \cite{159}, we have considered that the volume elements of the cosmic (DM) fluid perform isothermal flows (iDMF model). This assumption led us to an alternative approach to the DE concept. In fact, the internal energy of this fluid can compensate the (extra) DE needed for $\Om_0 = 1$ (\emph{cf.} Equation~(\ref{eq17})), even if the Universe is ever-decelerating (\emph{cf.} Equation~(\ref{eq25})). However, this is not the case for an observer who insists on treating DM as dust. To find out what is inferred by such an observer, we need to determine several cosmological parameters on which he/she relies, in interpreting observations. This can be done most appropriately by means of the conformal equivalence technique (\emph{cf.} Equations~(\ref{eq27}) and (\ref{eq28})), developed by Kleidis and Spyrou~\cite{175}. The~outcome is quite revealing.

In the iDMF model, there is a characteristic value of the actually-measured cosmological redshift, $z_c$ (\emph{cf.} Equation~(\ref{eq36})), above which, $d_L > \tilde{d}_L$. In other words, an observer who---although living in the iDMF model---treats DM as dust, infers that any SN Ia event located at $z > z_c$ lies farther (in other words, it looks fainter) than what is theoretically predicted (\emph{cf.} Equation~(\ref{eq35})). Furthermore, after the thermodynamical content of the iDMF model is taken into account, the theoretically-derived distance modulus, $\mu (z)$ (given by the combination of Equations~(\ref{eq35}), (\ref{eq37}) and (\ref{eq41})), fits the Hubble diagram of an extended sample of SNe Ia events \cite{181} quite accurately (green curve in Figure \ref{fig1}), in contrast to the corresponding collisionless-DM quantity, $\tilde{\mu} (\tilde{z})$ (dashed curve in Figure \ref{fig1}). In other words, in the iDMF model, no disagreement exists between observations and the theoretical prediction for the distant SNe Ia distribution. Finally, in the context of the iDMF approach, the observers who mistreat DM as dust, also infer that, recently (\emph{cf.} Equation~(\ref{eq48})), the Universe accelerated its expansion (\mbox{\emph{cf.} Equation~(\ref{eq47})). }

In view of all the above, the iDMF model could (most appropriately) serve as an alternative to the DE concept. Nevertheless, compatibility of this model with the observational data currently available, suggests that the DM consists of relativistic particles (\emph{cf.} Equation~(\ref{eq51})). In an effort to confront with the HDM issue, we have accordingly considered the (astrophysically) more relevant possibility that the volume elements of the DM fluid perform polytropic flows (pDMF model). 

The pDMF model is, simply, a conventional DE model \cite{160}. In the distant past, it behaves as a dust FRW model (\emph{cf.} Equation~(\ref{eq82}), for $z \gg 1$), while, on the approach to the present epoch, the internal physical characteristics of the cosmic fluid take over. In fact, at cosmological redshifts lower than $z = 0.384$, the energy density of the internal motions in the pDMF model dominates over its rest-mass counterpart (\emph{cf.} Equation~(\ref{eq69})). Once again, the internal energy compensates for the extra energy needed to compromise $\Om_0 = 1$ (\emph{cf.} Equation~(\ref{eq68})). In addition, for values of the polytropic exponent, $\Gm$, lower than unity, the pressure of the DM fluid is negative (\emph{cf.} Equation~(\ref{eq65})) and, so, the Universe accelerates its expansion at cosmological redshifts lower than a transition value (\emph{cf.} Equation~(\ref{eq84})), in a way also consistent with the condition $\varep + 3p < 0$ (\emph{cf.} Equations~(\ref{eq104}) and~(\ref{eq105})). 

Several physical requirements impose successive constraints on the value of the polytropic exponent. More specifically, the second law of Thermodynamics in an expanding Universe suggests that $\Gm \leq \gm$, where $\gm$ is the adiabatic index (in connection, see \cite{160}). In this context, for $\Gm < 1$, the pressure becomes negative (\emph{cf.} Equation~(\ref{eq65})). Furthermore, the condition for a non-negative transition redshift leads to $\Gm \leq 0.540$ (\emph{cf.} Equation~(\ref{eq85})), while, the requirement for a non-phantom Universe yields $\Gm > - 0.377$ (\emph{cf.} Equation~(\ref{eq72})). A positive velocity-of-sound square at all $z$, implies $\Gm \leq 0$ (\emph{cf.} Equation~(\ref{eq93})), and, eventually, the requirement for CDM at the present epoch results in $\Gm > -0.09$ (\emph{cf.} Equation~(\ref{eq97})). Hence, in a pDMF model that is compatible with modern observational cosmology, the polytropic exponent settles down to the range $-0.09 < \Gm \leq 0$, namely, if it is not zero, it is definitely negative, and very close to zero. 

In the pDMF approach, the theoretically-determined value of the deceleration parameter at the present epoch (Equation~(\ref{eq83})), has a well-shaped cross-section with the lower part of the corresponding observationally-determined range, $q_0 = -0.53_{-0.13}^{+0.17}$ \cite{218}. On the other hand, for $\Gm = 0$, the internal energy density becomes equal to its rest-mass counterpart at $z = 0.384$ (\emph{cf.} Equation~(\ref{eq69})), a theoretical prediction that lies well-within the corresponding $\Lm$CDM range, $z = 0.391 \pm 0.033$ \cite{30}. The pDMF approach can confront with every major cosmological issue, such as the age problem (see, e.g., Figure \ref{fig2}) and the coincidence problem (\emph{cf.} Equations~(\ref{eq69}) and (\ref{eq70}), as well as Equations~(\ref{eq104}) and (\ref{eq105})). What is most important, is that, in such a model, there is no disagreement between observations and the theoretical prediction of the distant SNe Ia distribution (see, e.g., Figure \ref{fig5}). Finally, along the lines of the pDMF framework, we can most naturally interpret why the Universe accelerates its expansion $z < z_{tr}$ (\emph{cf.} Equations~(\ref{eq104}) and (\ref{eq105})). 

In view of all the above, we conclude that, the cosmological model with matter content in the form of a self-interacting DM fluid performing either polytropic or isothermal flows, may serve either for a conventional DE model or a for viable alternative one, respectively. In fact, the idea that DE is nothing else but the shadowy reflection of DM, looks very promising and should be further inspected, in the search for a realistic approach to the DE concept. 

\acknowledgments{Acknowledgments}

Financial support by the Research Committee of the Technological Education Institute of Central Macedonia at Serres, Greece, under grant SAT/ME/211015-18/08, is gratefully acknowledged.

\authorcontributions{Author Contributions}

Both the authors of this article have substantially (and equally) contributed to the reported work.

\conflictofinterests{Conflicts of Interest}

The authors declare no conflict of interest.

\vspace{1.cm}

\bibliographystyle{mdpi}
\makeatletter
\renewcommand\@biblabel[1]{#1. }
\makeatother

\end{document}